\documentstyle[aps,pre,psfig]{revtex}

\begin{document}

\newcommand{\bp}{{\mbox{\boldmath$p$}}}
\newcommand{\bq}{{\mbox{\boldmath$q$}}}
\newcommand{\bx}{{\mbox{\boldmath$x$}}}
\newcommand{\by}{{\mbox{\boldmath$y$}}}
\newcommand{\bv}{{\mbox{\boldmath$v$}}}
\newcommand{\bA}{{\mbox{\boldmath$A$}}}
\newcommand{\bB}{{\mbox{\boldmath$B$}}}
\newcommand{\bC}{{\mbox{\boldmath$C$}}}
\newcommand{\bS}{{\mbox{\boldmath$S$}}}
\newcommand{\bK}{{\mbox{\boldmath$K$}}}
\newcommand{\bU}{{\mbox{\boldmath$U$}}}
\newcommand{\bTheta}{{\mbox{\boldmath$\Theta$}}}
\newcommand{\bPhi}{{\mbox{\boldmath$\Phi$}}}
\newcommand{\bLambda}{{\mbox{\boldmath$\Lambda$}}}
\newcommand{\bzeta}{{\mbox{\boldmath$\zeta$}}}
\newcommand{\bxi}{{\mbox{\boldmath$\xi$}}}

\newcommand{\ave}[1]{\left\langle#1\right\rangle}
\newcommand{\bave}[1]{\big\langle#1\big\rangle}
\newcommand{\Bave}[1]{\Big\langle#1\Big\rangle}
\newcommand{\dave}[1]{\langle\!\langle#1\rangle\!\rangle}
\newcommand{\bdave}[1]{\big\langle\!\!\big\langle#1\big\rangle\!\!\big\rangle}
\newcommand{\Bdave}[1]{\Big\langle\!\!\Big\langle#1\Big\rangle\!\!\Big\rangle}

\wideabs{
\title{Thermodynamics of quantum dissipative many-body systems}
\author{Alessandro Cuccoli\cite{e-AC}, Andrea Fubini\cite{e-AF},
        Valerio Tognetti\cite{e-VT}}
\address{Dipartimento di Fisica dell'Universit\`a di Firenze
    and Istituto Nazionale di Fisica della Materia (INFM),
    \\ Largo E. Fermi~2, I-50125 Firenze, Italy}
\author{Ruggero Vaia\cite{e-RV}}
\address{Istituto di Elettronica Quantistica
    del Consiglio Nazionale delle Ricerche,
    via Panciatichi~56/30, I-50127 Firenze, Italy,
    \\ and Istituto Nazionale di Fisica della Materia (INFM)}
\date{\today}
\maketitle

\begin{abstract}
We consider quantum nonlinear many-body systems with dissipation described
within the Caldeira-Leggett model, i.e., by a nonlocal action in the path
integral for the density matrix.
Approximate classical-like formulas for thermodynamic quantities are derived
for the case of many degrees of freedom, with general kinetic and dissipative
quadratic forms.
The underlying scheme is the {\em pure-quantum self-consistent harmonic
approximation} (PQSCHA), equivalent to the variational approach by the
Feynman-Jensen inequality with a suitable quadratic nonlocal trial action.
A low-coupling approximation permits to get manageable PQSCHA expressions
for quantum thermal averages with a classical Boltzmann factor involving an
effective potential and an inner Gaussian average that describes
the fluctuations originating from the interplay of quanticity
and dissipation.
The application of the PQSCHA to a quantum $\phi^4$-chain with Drude-like
dissipation shows nontrivial effects of dissipation, depending upon its
strength and bandwidth.
\end{abstract}


} 

\section{Introduction}

Historically, the classical effective potential was introduced in 1955 by
Feynman~\cite{Feynman55} for treating the polaron, which can be regarded
as an electron subjected to a ``dissipative'' interaction with the lattice
phonons.
A remarkable improvement of Feynman's effective potential in the nondissipative
case~\cite{FeynmanH65,Feynman72} was obtained~\cite{GT85prl,FeynmanK86} 3
decades later by introducing a quadratic term and an extra variational
parameter in the trial action used to approximate the partition function.

Several applications of the improved method to condensed matter systems have
demonstrated its usefulness (see~\cite{CGTVV95,Kleinert95} and references
therein).
The fact that it is also suitable to treat open systems was used in previous
studies~\cite{KimC90,FalciFG91} basically aimed to obtain
a classical-like expression for the free energy in the case of strictly
linear dissipation~\cite{Weiss99}; the case of nonlinear dissipation was
also considered~\cite{BaoZW95}.

In a previous paper~\cite{CRTV97} we used the method, also called the
{\em pure-quantum self-consistent harmonic approximation} (PQSCHA) after its
generalization to phase-space Hamiltonians~\cite{CTVV92ham,CGTVV95}, to
obtain the density matrix of a single particle with nonlinear interaction
and with dissipation described through the Caldeira-Leggett (CL)
model~\cite{CaldeiraL83,Weiss99}.
In the present work we deal with many degrees of freedom, facing the
problem of making the method suitable for actual applications to
condensed matter systems.

In Section~\ref{s.PQSCHA} we develop the general method, as well as the
necessary low-coupling approximation and its specialization to the most
common case of translation invariant systems.
In Section~\ref{s.phi4} the kink-bearing quantum $\phi^4$ chain is
considered and the corresponding effective classical potential is
explicitly derived. Finally, results for some thermodynamic quantities of
this system are collected and illustrated in Section~\ref{s.results}.

\section{PQSCHA for dissipative many-body systems}
\label{s.PQSCHA}

\subsection{Path integral for the density matrix}
\label{ss.pathint}

Let us consider a general system with $N$ degrees of freedom, i.e., canonical
coordinate and momentum operators $\hat\bq\equiv \{\hat{q}_i \}_{i=1,...,N}$
and $\hat\bp\equiv \{\hat{p}_i \}_{i=1,...,N}$, with the commutation relations
$[\hat{q}_i,\hat{p}_j]= {\rm i}\delta_{ij}$ (we put $\hbar=1$ from now on),
and described by a Hamiltonian with a quadratic kinetic energy and a nonlinear
potential term,
\begin{equation}
\label{e.hatH}
\hat{\cal{H}} = \frac12\, {}^{\rm{t}}\!\hat\bp\, \bA^2 \hat\bp + V(\hat\bq)~.
\end{equation}
The matrix $\bA^2=\{A_{ij}^2 \}$ is symmetric, real, and positive definite;
thus its positive square root $\bA$ and its inverse $\bA^{-1}$ do exist;
it is convenient to define its determinant in terms of a positive `mass' $m$,
\begin{equation}
 \det\bA^2 \equiv m^{-N} ~.
\label{e.m}
\end{equation}

In order to introduce dissipation in this system, further information is
needed about the physics of the dissipation mechanism.
We assume this to be described through the CL model, i.e., introducing
an environment (or damping bath) of harmonic degrees of freedom, but still
one can think of different kinds of environmental coupling.
For instance, several damping baths coupled with the coordinates can be
used to describe different dissipation mechanisms, and it can be even
necessary to introduce infinite (correlated or uncorrelated) baths;
since there are many `particles' in the system, it may happen that all
particles are coupled with one single environment, so introducing correlations
via dissipation, or -- in the simplest physical model -- that there are
infinite identical independent baths, one for each particle.
We just mention here that -- although here we stick with coordinate coupling
-- it is also possible to couple the environment with
momenta~\cite{ZwergerDF86}, or with both coordinates and momenta: the framework
we are going to describe can be also generalized to these cases.

The density matrix at the equilibrium temperature $T=\beta^{-1}$ in the
coordinate representation is expressed by Feynman's path integral as
\begin{equation}
 \rho(\bq'',\bq')\equiv\langle\bq''|e^{-\beta\hat{\cal H}_{\rm tot}}|\bq'\rangle
 =\int_{\bq'}^{\bq''}{\cal{D}}\big[\bq(u)\big]~e^{-S[\bq(u)]}~,
\label{e.rhoqq}
\end{equation}
where $\hat{\cal{H}}_{\rm{tot}}$ is the sum of~(\ref{e.hatH}) and the
Hamiltonians for the bath and the system-bath coupling, and the average
with respect to the bath variables (usually called `tracing out') is
understood~\cite{Weiss99}.
The path integration is defined as a sum over all paths $\bq(u)$,
with $u\in[0,\beta]$, $\bq(0)=\bq'$, and $\bq(\beta)=\bq''$.

The general CL Euclidean action for the system, obtained after having
traced out all environmental variables, takes the form
\begin{eqnarray}
 & & S\big[\bq(u)\big] = \int_0^\beta du \bigg[
 \frac12\, {}^{\rm{t}}\!\dot\bq(u){\mbox{\boldmath$A$}}^{-2}\dot\bq(u)+V\big(\bq(u)\big)
\nonumber\\
 & & \hspace{4mm}
 - \int_0^\beta \! \frac{du'}4\,{}^{\rm{t}}\!\big(\bq(u){-}\bq(u')\big)
 \bK(u{-}u')\,\big(\bq(u){-}\bq(u')\big) \!\bigg] \,,
 \label{e.S}
\end{eqnarray}
where the kernel matrix $\bK(u)=\{K_{ij}(u)\}$ is a real symmetric matrix that
replaces the scalar kernel $k(u)$ of the single-particle case~\cite{Weiss99};
as a function of $u$ it keeps its symmetry and periodicity,
$\bK(u)=\bK(-u)=\bK(\beta-u)$, and satisfies $\int_0^\beta{du}\,\bK(u)=0$.
In the case of $N$ independent identical environmental baths coupled to each
coordinate $\hat{q}_i$ one simply has a diagonal kernel:
\begin{equation}
 \bK(u)=k(u)~.
\label{e.Kdiag}
\end{equation}
We will consider this case for the application shown in Section~\ref{s.phi4}.

\subsection{General PQSCHA}
\label{ss.PQSCHA}

As suggested by Feynman \cite{Feynman72}, the path integral (\ref{e.S})
can be rearranged by summing over classes of paths that share the same
{\em average point} $\bar\bq$\,:
\begin{equation}
 \rho(\bq'',\bq') =\int d^N\bar{\bq}~ \bar\rho(\bq'',\bq';\bar{\bq})~,
\label{e.rhobrho}
\end{equation}
\begin{equation}
 \bar\rho(\bq'',\bq';\bar\bq) = \int_{\bq'}^{\bq''}
 \!\!\! {\cal{D}}\big[\bq(u)\big]~\delta\Big(\bar\bq{-}\bar\bq\big[\bq(u)\big]\Big)
 ~e^{-S[\bq(u)]}~,
\label{e.brho}
\end{equation}
where $\bar\bq\big[\bq(u)\big]=(\beta)^{-1}\int_0^\beta{du}\,\bq(u)$ is the
average-point functional.
The PQSCHA is based on approximating the action $S\big[\bq(u)\big]$ with a trial
action $S_0\big[\bq(u)\big]$ that is quadratic so that the path integral can be
evaluated analytically, and contains parameters that can be optimized.
It is to be remarked that only paths with a fixed $\bar\bq$ appear into
the path integral~(\ref{e.brho}), so that $S_0\big[\bq(u)\big]$ (i.e., the
parameters appearing therein) can depend on $\bar\bq$: one deals then with a
much more general class of trial actions than one could obtain taking sensible
approximations of the Hamiltonian operator -- as for instance in the usual
self-consistent harmonic approximation (SCHA) -- , and a better approximation
is then to be expected; the price is that the classical-like integral over
$\bar\bq$ is left over. We define then the trial action by replacing $V(\bq)$
in the action~(\ref{e.S}) with a trial quadratic `potential',
\begin{equation}
 V_0(\bq;\bar\bq) = w(\bar\bq)
 +\frac12\, {}^{\rm{t}}\!(\bq{-}\bar\bq)\,\bB^2(\bar\bq)\,(\bq{-}\bar\bq)~.
\label{e.V0}
\end{equation}
The parameters are the scalar $w(\bar\bq)$ and the $N(N{+}1)/2$
components of the symmetric real matrix $\bB^2(\bar\bq)$.
These are to be optimized in such a way that the trial reduced density
$\bar\rho_0(\bq'',\bq';\bar\bq)$ at best approximates
$\bar\rho(\bq'',\bq';\bar\bq)$, {\it for each value of} $\bar\bq$.

The off-diagonal elements of the trial reduced density are rather tricky
to evaluate, since the general method of calculating the minimal action
cannot be used (the classical path being the solution of infinite-order equations
of motion), while the method of Fourier expansion of the paths can be applied
only to integrals over closed paths (but $\bq''\neq\bq'$ for the off-diagonal
part of the density matrix).
However, the latter method may still be used if one closes the paths in a
small interval and separately evaluates the corresponding contribution.
Explicitly:
\begin{eqnarray}
 & & \bar\rho_0(\bq'',\bq';\bar\bq) =
 \lim\limits_{\varepsilon\to 0} ~\frac{1}{{\cal F}_\varepsilon}
 \oint {\cal{D}}\big[\bq(u)\big] ~e^{-S_0[\bq(u)]}
\nonumber\\
 & & ~~~~~~~~\times
 \,\delta\big(\bar\bq{-}\bar\bq\big[q(u)\big]\big)
 \,\delta\big(\bq(\varepsilon){-}\bq'\big)
 \,\delta\big(\bq(\beta{-}\varepsilon){-}\bq''\big) ~;
\label{e.brho0eps}
\end{eqnarray}
here the integral is over all {\it closed} paths
$\{\bq(u)\,|\,u\in[0,\beta]\,\}$ satisfying the constraints
$\bq(\varepsilon)=\bq'$ and $\bq(\beta-\varepsilon)=\bq''$;
${\cal{F}}_\varepsilon$ is the integral over the open paths
$\{\,\bq(u)\,|\,u\in[-\varepsilon,\varepsilon]\,\}$ (the range
$[\beta-\varepsilon,\beta]$ is periodically mapped onto
$[-\varepsilon,0]$) with end points $\bq(-\varepsilon)=\bq''$ and
$\bq(\varepsilon)=\bq'$. In the limit of small $\varepsilon$,
${\cal{F}}_\varepsilon$ becomes the free-particle density matrix,
\begin{equation}
 {\cal F}_\varepsilon = \bigg(\frac{m}{4\pi\varepsilon}\bigg)^{N/2}
 ~e^{-{}^{\rm{t}}\!\bzeta\,\bA^{-2}\bzeta/(4\varepsilon)
 +O(\varepsilon)}\,.
\label{e.F-eps}
\end{equation}
where $\bzeta\equiv\bq''-\bq'$.
The paths in Eq.~(\ref{e.brho0eps}) can thus be Fourier expanded,
\begin{equation}
 \bq(u)=\bar\bq+2\sum\limits_{n=1}^{\infty} ~(\bx_n\cos\nu_nu+\by_n\sin\nu_nu)~;
\label{e.qn}
\end{equation}
$\nu_n=2\pi{n}/\beta$ are the Matsubara frequencies, the $0^{\rm{th}}$
component is just $\bar\bq$. The measure of the path integral
becomes~\cite{Feynman72}
\begin{equation}
 \oint{\cal{D}}\big[\bq(u)\big] \to
 \bigg(\frac{m}{2\pi\beta}\bigg)^{\!\textstyle{N\over2}} \!\!\!
 \int \! d\bar\bq\,\prod\limits_{n=1}^{\infty} \!
 \bigg(\frac{m\beta\nu_n^2}{\pi}\bigg)^N \!\!\! \int d\bx_nd\by_n~,
\label{e.measure}
\end{equation}
and the trial action takes the form
\begin{equation}
 S_0\big[\bq(u)\big] = \beta w(\bar\bq)
 +\beta\sum\limits_{n=1}^{\infty} \Big[\,
 {}^{\rm{t}}\!\bx_n \bPhi_n \bx_n +{}^{\rm{t}}\!\by_n \bPhi_n \by_n \Big]~.
\label{e.S0Matsu}
\end{equation}
where the matrices $\bPhi_n=\{\Phi_{n,ij}\}$ are given by
\begin{equation}
 \bPhi_n(\bar\bq)= \nu_n^2\bA^{-2} + \bB^2(\bar\bq) + \bK_n
\label{e.Phi}
\end{equation}
and $\bK_n$ is the Matsubara transform of the kernel matrix $\bK(u)$,
\begin{equation}
 \bK_n= \int_0^\beta du~\bK(u)~\cos\nu_nu~.
\end{equation}
The calculation of the reduced density~(\ref{e.brho0eps}) is reported in
Appendix~\ref{a.rho0}; furthermore, it is apparent that no confusion
arises if we {\em suppress the bar} from $\bar\bq$ in the rest of the paper.

The simplest way for writing the final result is to give the expression
of the thermal average of a generic observable $\hat{\cal{O}}$ in terms of
its Weyl symbol~\cite{Berezin80,HilleryCSW84,CTVV92ham}, whose definition is
\begin{equation}
 {\cal{O}}(\bp,\bq)=\int d\bzeta~e^{i\,{}^{\rm{t}}\!\bp\,\bzeta}~
 \big\langle\bq{+}{\textstyle\frac12}\bzeta\big|\hat{O}
 \big|\bq{-}{\textstyle\frac12}\bzeta\big\rangle ~.
\label{e.Weylsymb}
\end{equation}
Indeed, using the trace property of Weyl symbols,
\begin{equation}
 {\rm{Tr}}\,\hat\rho\,\hat{\cal{O}}
 =\int \frac{d\bp\,d\bq}{(2\pi)^N}~\rho(\bp,\bq)\,{\cal{O}}(\bp,\bq) ~,
\label{e.aveWeyl}
\end{equation}
one finds the fundamental formula that approximates quantum averages by
means of a classical-like expression with an effective potential $V_{\rm{eff}}$,
\begin{equation}
 \bave{\hat{\cal{O}}}=\frac1{{\cal{Z}}} \bigg(\frac{m}{2\pi\beta}\bigg)^{N/2}
 \int d\bq ~e^{-\beta V_{\rm{eff}}(\bq)}
 \bdave{{\cal{O}}(\bp,\bq{+}\bxi)} ~,
\label{e.aveO}
\end{equation}
where $\dave{\,\cdot\,}$ is the Gaussian average over the variables
$\bp$ and $\bxi$ determined by $\bar\rho_0$ as reported in Eq.~(\ref{e.brho0pq});
this average can be uniquely defined through its nonzero moments
$\bdave{\bxi\,{}^{\rm{t}}\!\bxi}=\bC(\bq)$ and
$\bdave{\bp\,{}^{\rm{t}}\!\bp}=\bLambda(\bq)$, with
\begin{eqnarray}
 \bC(\bq) &=&  \frac2\beta\,\sum_{n=1}^\infty \bPhi_n^{-1}(\bq) ~,
\label{e.Cij}
\\
 \bLambda(\bq) &=& \frac1\beta\sum_{n=-\infty}^\infty
 \bA^{-2}\,\big[\bA^2- \nu_n^2\bPhi_n^{-1}(\bq)\big]\,\bA^{-2} ~,
\label{e.Lambdaij}
\end{eqnarray}
whose components are the {\em renormalization coefficients};
the effective potential reads
\begin{equation}
 V_{\rm{eff}}(\bq) = w(\bq) + \beta^{-1} \mu(\bq) ~,
\label{e.Veff}
\end{equation}
with
\begin{equation}
 \mu(\bq) = \sum\limits_{n=1}^{\infty}\,
 \ln\,\frac{\det\bPhi_n(\bq)}{(m\nu_n^2)^N}~,
\label{e.mumatr}
\end{equation}

To implement the PQSCHA we require~\cite{CTVV92ham,CGTVV95} that the parameters
of the trial action are such to match the $\rho_0$-averages of the original
and the trial potential and of their second derivatives:
\begin{eqnarray}
 w(\bq)&=&\bdave{V(\bq+\bxi)}
 -{\textstyle\frac12}\,{\rm{Tr}}\,\big[\bB^2(\bq)\,\bC(\bq)\big] ~,
\label{e.w}
\\
 B^2_{ij}(\bq)&=&\bdave{\partial_{q_i}\partial_{q_j}V(\bq+\bxi)} ~.
\label{e.Bij}
\end{eqnarray}
The second equation together with Eq.~(\ref{e.Cij}) self-consis\-tently
determines the solution for $\bB(\bq)$ and $\bC(\bq)$.

\subsection{Low-coupling approximation}
\label{ss.LCA}

The above framework is still complicated, due to the dependence on $\bq$
that requires a solution of Eq.~(\ref{e.Bij}) for any value of $\bq$, and
also due to the matrix form of Eq.~(\ref{e.Cij}).
The first difficulty can be overcome by the so-called {\em low-coupling
approximation} (LCA) and consists in expanding the $N{\times}N$ matrix
$\bB(\bq)$ so to make the renormalization coefficients, and hence also the
Gaussian averages $\dave{\,\cdot\,}$, independent of the configuration $\bq$.

In order to do this it is useful to introduce the differential operator
\begin{equation}
 \Delta(\bq)=\frac12\sum_{ij}C_{ij}(\bq)~\partial_{q_i}\partial_{q_j}
\end{equation}
so that one can write, for any function $F(\bq)$,
\begin{equation}
 \bdave{F(\bq+\bxi)}=e^{\Delta(\bq)}~F(\bq) ~,
\end{equation}
where the derivatives are assumed {\em not} to operate on the renormalization
coefficients $C_{ij}(\bq)$. In view of Eq.~(\ref{e.w})
this allows us to express the effective potential~(\ref{e.Veff}) as
\begin{equation}
 V_{\rm{eff}}(\bq) = \big[1-\Delta(\bq)\big]e^{\Delta(\bq)}\,V(\bq)
 + \beta^{-1}\,\mu(\bq) ~.
\label{e.VeffDelta}
\end{equation}
Expanding from the configuration $\bq_0$ that minimizes
$V_{\rm{eff}}(\bq)$, i.e., setting $\bB^2(\bq)=\bB^2+\delta\bB^2(\bq)$
with the convention of dropping the fixed argument $\bq_0$, i.e.,
$\bB\equiv\bB(\bq_0)$, from the definition~(\ref{e.Phi}) one has
\begin{eqnarray}
 \det\bPhi_n(\bq) &=& \det\big[\bPhi_n+\delta\bB^2(\bq)\big]
\nonumber\\
 &\simeq& \det\bPhi_n\big\{1+{\rm{Tr}}\big[\bPhi_n^{-1}\delta\bB^2(\bq)\big]\big\}
\end{eqnarray}
and for the last term of the effective potential~(\ref{e.Veff})
this gives
\begin{eqnarray}
 \beta^{-1}\,\mu(\bq) &\simeq& \beta^{-1}\,\mu
 +{\textstyle\frac12}{\rm{Tr}}\big[\bC\,\delta\bB^2(\bq)\big]
\nonumber\\
 & & ~= \beta^{-1}\,\mu + \Delta \big[e^{\Delta(\bq)}V(\bq)- e^\Delta V(\bq_0)\big]\,,
\end{eqnarray}
so that Eq.~(\ref{e.VeffDelta}) becomes the LCA effective potential,
\begin{equation}
 V_{\rm{eff}}(\bq) = e^\Delta\,V(\bq) -\Delta e^\Delta V(\bq_0) + \beta^{-1}\,\mu ~,
\label{e.VeffLCA}
\end{equation}
where the whole dependence on $\bq$ is contained in the first term
$e^\Delta\,V(\bq)=\bdave{V(\bq{+}\bxi)}$, calculated with the renormalization
coefficients independent of the configuration, $\bC=\bC(\bq_0)$. It is useful
to note that for this reason one can simply write, from Eq.~(\ref{e.Bij}),
\begin{equation}
 B^2_{ij}=\partial_{q_i}\partial_{q_j}\,V_{\rm{eff}}(\bq_0)~.
\label{e.BijLCA}
\end{equation}

It often occurs that the indices $i,j,\dots$ refer to the sites of
a lattice, whose symmetries can be very helpful in order to
simplify the analysis, provided that the minimum configuration of
$V_{\rm{eff}}(\bq)$ shares the same property.
In general $\bB^2(\bq)$ has no particular symmetries, due to its
dependence on $\bq$: but the LCA matrix $\bB^2=\bB^2(\bq_0)$ is much
likely to have the same symmetries of the Hamiltonian, provided that
the minimum configuration $\bq_0$ shares them.

Let us consider the most frequent case of translation symmetry: if the
Hamiltonian and the dissipation kernel $\bK(u)$ are translation invariant,
the calculations are greatly simplified. Then all the matrices are
diagonalized (apart from internal degrees of freedom, which we do not
consider here) by an (orthogonal) Fourier transform $\bU=\big\{U_{ki}\big\}$\,:
\begin{eqnarray}
 m^{-1}_k~\delta_{kk'}&=&{\sum}_{ij}~U_{ki}\,U_{k'j}\,A^2_{ij} ~,
\label{e.Ak}
\\
 m_k\,\omega^2_k~\delta_{kk'}&=&{\sum}_{ij}~U_{ki}\,U_{k'j}\,B^2_{ij} ~,
\label{e.Bk}
\\
 m_kK_{n,k}~\delta_{kk'}&=&{\sum}_{ij}~U_{ki}\,U_{k'j}\,K_{n,ij} ~,
\label{e.Kk}
\end{eqnarray}
where we used `familiar' notations in terms of masses and frequencies,
in order to compare with the known expressions for one single degree of
freedom~\cite{CRTV97}. It is immediately seen from Eq.~(\ref{e.m}) that
$\prod_km_k=m^N$\,.
Taking the LCA matrix $\bPhi_n=\nu_n^2\bA^{-2}+\bB^2+\bK_n$ the last term of
the effective potential entails the quantity
\begin{equation}
 \mu=\sum_k\,\sum_{n=1}^\infty\ln\frac{\nu_n^2+\omega^2_k+K_{n,k}}{\nu_n^2} ~.
\label{e.muLCA}
\end{equation}
The renormalization coefficients of Eqs.~(\ref{e.Cij}) and~(\ref{e.Lambdaij})
become, for the Fourier transformed variables $\xi_k=\sum_iU_{ki}\xi_i$ and
$p_k=\sum_iU_{ki}p_i$\,,
\begin{eqnarray}
 C_{kk'} = \bdave{\xi_k\xi_{k'}} &=&
 \delta_{kk'}~ \frac{2}{\beta m_k}\,
 \sum_{n=1}^\infty \frac1{\nu_n^2+\omega^2_k+K_{n,k}} ~,
\label{e.CkLCA}
\\
 \Lambda_{kk'} = \bdave{p_k p_{k'}} &=&
 \delta_{kk'}~ \frac{m_k}\beta\sum_{n=-\infty}^\infty
 \frac{\omega^2_k+K_{n,k}}{\nu_n^2+\omega^2_k+K_{n,k}} ~.
\label{e.LambdakLCA}
\end{eqnarray}
Their counterparts in direct space are thus easily recovered: for instance, the
`on-site' renormalization coefficient $D\equiv C_{ii}$ can be simply expressed as
\begin{equation}
 D \equiv \bdave{\xi_i^2} = \frac{2}{\beta N} \sum_k \frac1{m_k}\,
 \sum_{n=1}^\infty \frac1{\nu_n^2+\omega^2_k+K_{n,k}} ~.
\label{e.D}
\end{equation}
We remark that the renormalization coefficients
$C_{ij}=\bdave{\xi_i\xi_j}$ describe not only the pure quantum
fluctuations~\cite{CGTVV95} but also those arising from the dissipative
coupling; on the other hand, the $\Lambda_{ij}=\bdave{p_ip_j}$ includes
also the classical fluctuations -- the sum in~(\ref{e.LambdakLCA}) contains
indeed the $n=0$ term -- since for a standard Hamiltonian as~(\ref{e.hatH})
they are Gaussian and there is no reason for keeping them separated.

The partition function and thermal averages are then to be evaluated by
means of Eq.~(\ref{e.aveO}), where, of course, the effective potential and the
double-bracket average are to be understood as the LCA ones.
Note that, since the LCA $\Lambda$'s do not depend on $\bq$, averages
of observables involving only the momenta are trivially evaluated as
$\bave{{\cal{O}}(\hat\bp)}=\bdave{{\cal{O}}(\bp)}$.
In the next Section we illustrate the method by applying the above results
to a model nonlinear many-body system.

\section{Dissipative $\phi^4$ chain}
\label{s.phi4}

\subsection{The model}

The nondissipative $\phi^4$ chain has been already
studied~\cite{GTV88fi4,CGTVV95} by the effective potential method.
The model consists of a one-dimensional array of particles with a
nearest-neighbor harmonic interaction and a quartic on-site interaction.
It can be viewed as the discretized version of a continuum nonlinear
field theory, described by the (undamped) action
\begin{equation}
 S= \frac1g \int_0^\beta du~a\,\sum_i\bigg[\frac{\,\dot q_i^2}2
 +\frac{(q_i-q_{i-1})^2}{2a^2} + \Omega^2 v(q_i) \bigg]~,
\label{e.Sphi4}
\end{equation}
where $a$ is the chain spacing, $\Omega$ is the gap of the bare dispersion
relation, $g$ is the quantum coupling, and $v(q_i)$ is the local nonlinear
potential,
\begin{equation}
 v(q)=\frac18\, \big(1-q^2\big)^2 ~.
\label{e.vphi4}
\end{equation}
which is symmetric and has two wells in $q_0=\pm{1}$ with $v''(q_0)=1$
separated by a barrier. The index $i=1,...,N$ and periodic boundary
conditions ensure the translation symmetry.

The classical system has two degenerate translation-invariant absolute
minimum configurations, $\{q_i=1\}$ and $\{q_i=-1\}$, as well as relative
minima, the static `kinks', with the configuration going over from one well
to the other (e.g., $q_i\to\pm{1}$ for $i\to\pm\infty$, respectively).
In the continuum limit $ia\to{x}$ the general kink configuration is indeed
$q(x)=\pm\tanh\Omega(x-x_0)$, so that it is localized with a characteristic
length $\Omega^{-1}$ (the `relativistic' velocity has been set to 1 in
Eq.~(\ref{e.Sphi4}), so that length and time are dimensionally equivalent),
and its energy is $\varepsilon_{{}_{\rm{K}}}=2\Omega/3g$.

The quantum behavior of the system is ruled by the ratio between the
characteristic frequency of the quasi harmonic excitations of the system
$\Omega$ and the energy scale $\varepsilon_{{}_{\rm{K}}}$, so that we are
lead to introduce the coupling parameter
\begin{equation}
 Q=\frac{\Omega}{~\varepsilon_{{}_{\rm{K}}}}=\frac{\,3\,}2~g ~,
\end{equation}
that we will use in the place of $g$, using the notations of
Ref.~\cite{GTV88fi4}; another useful dimensionless parameter is the kink
length in lattice units, $R=1/a\Omega$ ($R\to\infty$ in the continuum limit).
It turns out to be useful to use $\Omega$ as the natural frequency scale,
and also $\varepsilon_{{}_{\rm{K}}}=\Omega/Q$ as the overall energy scale,
since the most interesting thermodynamic features are
displayed when kinks are thermally excited in the system; for instance,
this results in a peak of the specific heat at $t\sim{0.2}$, where
$t\equiv{T}/\varepsilon_{{}_{\rm{K}}}$ is the reduced temperature
used from now on.

The $\phi^4$ chain Hamiltonian can then be written as
\begin{eqnarray}
 \hat{\cal H}&=& \varepsilon_{{}_{\rm{K}}} \bigg[\,
 \frac{Q^2R}3\sum_i\hat p_i^2+V(\hat\bq)\, \bigg]~,
\label{e.Hphi4}
\\
 V(\bq)&=& \frac{3}{2R}\sum_i\bigg[v(q_i)
 +\frac{\,R^2}2\,(q_i-q_{i-1})^2\bigg] ~,
\label{e.Vphi4}
\end{eqnarray}
and from Eq.~(\ref{e.Ak}) one can immediately find the
identification $m_k^{-1}=m^{-1}=2Q^2R\varepsilon_{{}_{\rm{K}}}/3$.
The meaning of $Q$ as `quantum coupling' becomes
immediately transparent if one thinks to absorb it into $\hat\bp$, such that
$[\bq_i,\bp_j]=i\,Q\,\delta_{ij}$, where $Q$ plays the role of $\hbar$, indeed.

To introduce dissipation in the $\phi^4$ chain, we use the simplest CL model
with $N$ independent environmental baths, that gives a diagonal kernel matrix,
Eq.~(\ref{e.Kdiag}).
In order to make contact with the usual formalism, we write the dissipation
kernel as a Fourier transform,
\begin{equation}
 k(u) = {m\over\beta} \sum\limits_{n=-\infty}^{\infty} e^{i\nu_n u} ~k_n~,
\end{equation}
where $\nu_n=2\pi{n}/\beta=2\pi{n}\Omega{t}/Q$ are the Matsubara frequencies,
setting then~\cite{Weiss99}
\begin{equation}
 k_n\equiv K_{n,k}=~|\nu_n|~\gamma\big(z=|\nu_n|\big)~,
\label{e.kn}
\end{equation}
where $\gamma(z)$ is the Laplace transform of the real-time memory
damping function $\gamma(t)$ that would appear in the Langevin equation
derived by elimination of the bath variables the CL model equations of
motion~\cite{Weiss99}.

We shall use a {\em Drude}-like spectral function of the environmental
interaction~\cite{Weiss99}, i.e.,
\begin{equation}
 \gamma(z)=\frac{\gamma\,\omega_{{}_{\rm{D}}}}{\omega_{{}_{\rm{D}}}+z} ~,
\label{e.gammaDz}
\end{equation}
where the constant $\gamma$ rules the {\em strength} of the coupling with the
dissipation bath, while the Drude frequency $\omega_{{}_{\rm{D}}}$ characterizes
its `{\em bandwidth}'. For instance, a given degree of freedom is expected not to
interact with the dissipation bath (and thus not to dissipate) if its
characteristic frequency is much larger than $\omega_{{}_{\rm{D}}}$.

Eq.~(\ref{e.gammaDz}) corresponds to taking a (retarded) real-time memory
damping function with exponential decay
\begin{equation}
 \gamma(t)=\theta(t)~\gamma\,\omega_{{}_{\rm{D}}}~e^{-\omega_{{}_{\rm{D}}}t} ~.
\label{e.gammaDt}
\end{equation}
This becomes a $\delta$ function in the {\em Ohmic} (or {\em Markovian}) limit
$\omega_{{}_{\rm{D}}}\to\infty$\,:
\begin{equation}
 \gamma(z)=\gamma ~,~~~~~~~~ \gamma(t)=\gamma~\delta(t-0^+)~.
\label{e.gammaOhmic}
\end{equation}

From Eqs.~(\ref{e.kn}) and~(\ref{e.gammaDz}) and introducing the reduced damping
strength and Drude frequency
\begin{equation}
 \Gamma=\gamma/\Omega ~,~~~\Omega_{{}_{\rm{D}}}=\omega_{{}_{\rm{D}}}/\Omega
\end{equation}
the dissipation kernel, which has the dimension of a square frequency, becomes
\begin{equation}
 \frac{k_n}{\Omega^2}=
 \Gamma\Omega_{{}_{\rm{D}}} \frac{\nu_n/\Omega}{\Omega_{{}_{\rm{D}}}+\nu_n/\Omega}~.
\label{e.kn1}
\end{equation}

\subsection{Effective potential}
\label{ss.phi4.veff}

As already remarked, the relevant nonlinear effects on the thermodynamic
behavior of the $\phi^4$ chain occur at finite temperature, when kinks are
excited in the system. Therefore, in the study of the quantum system it is
crucial to retain the overall nonlinearity that gives rise to the kink
solutions. This goal cannot be achieved by a perturbative
approach~\cite{MakiT79} and the effective potential method appears the only
one through which one can obtain significant results.

The recipe~(\ref{e.VeffLCA}) of the previous Section for deriving the (LCA)
effective potential applies very simply to the potential~(\ref{e.Vphi4}), once
one assumes to take a translation invariant minimum $\bq_0=\{q_{0,i}{=}q_0\}$.
Since any quadratic terms, as the nearest-neighbor term in~(\ref{e.Vphi4}),
remain unchanged, it appears that the effective potential can be written in
the same form of Eq.~(\ref{e.Vphi4}),
\begin{equation}
 V_{\rm{eff}}(\bq) = \frac{3}{2R}\sum_i\bigg[v_{\rm{eff}}(q_i)
 +\frac{\,R^2}2\,(q_i-q_{i-1})^2\bigg] ~,
\label{e.Veffphi4}
\end{equation}
where $v(q)$ is replaced by an effective local potential,
\begin{equation}
 v_{\rm{eff}}(q)=\frac18\big[q^2-1+3D(t)\big]^2+\frac34D^2(t)
 +t \tilde\mu(t) \,.
\label{e.veffphi4}
\end{equation}
$D(t)=D(t;Q,R,\Gamma,\Omega_{{}_{\rm{D}}})$ is the renormalization coefficient
defined in Eq.~(\ref{e.D}) and $\tilde\mu(t)=(2R/3N)\mu(t)$, with $\mu(t)$ as
defined in Eq.~(\ref{e.muLCA}).
The (doubly degenerate) minimum is in $q_0=\pm\sqrt{1{-}3D}$, i.e., the wells
are `effectively' closer by the effect of quantum fluctuations.

From Eq.~(\ref{e.BijLCA}) one finds
\begin{equation}
 B_{ij}=\frac{3\varepsilon_{{}_{\rm{K}}}}{2R}\big[v_{\rm{eff}}''(q_0)\,\delta_{ij}
 +R^2(2\delta_{ij}{-}\delta_{i,j+1}{-}\delta_{i,j-1})\big]
\end{equation}
and, being $v_{\rm{eff}}''(q_0)=1-3D$ its Fourier transform~(\ref{e.Bk}) gives
the renormalized frequencies
\begin{equation}
 \frac{\omega_k^2}{\Omega^2}\equiv\Omega_k^2 =1-3D+4R^2\sin^2\frac{k}2 ~.
\label{e.omegakphi4}
\end{equation}
The dimensionless wavevector $k$ takes $N$ values in $[-\pi,\pi]$ and
$D(t)$ and $\tilde\mu(t)$ can be rewritten as
\begin{eqnarray}
 D(t)&=&\frac{Q^2R}{3tN}\sum_k\sum_{n=1}^\infty
 \frac{1}{(\pi n)^2+f_k^2+\tilde{k}_n} ~,
\label{e.Dphi4}
\\
 \tilde\mu(t)&=&\frac{2R}{3N}\sum_k\sum_{n=1}^\infty
 \ln\,\frac{(\pi n)^2+f_k^2+\tilde{k}_n}{(\pi n)^2} ~,
\label{e.muphi4}
\end{eqnarray}
where  $f_k=Q\Omega_k/(2t)$, and $\tilde{k}_n$ can be derived from
Eqs.~(\ref{e.kn1}),
\begin{equation}
 \tilde k_n=\frac{Q\Gamma}{2t}\,\frac{f_{{}_{\rm{D}}}\,\pi n}{f_{{}_{\rm{D}}}+\pi n} ~,
\label{e.knphi4}
\end{equation}
with $f_{{}_{\rm{D}}}=Q\Omega_{{}_{\rm{D}}}/(2t)$.
Note that Eqs.~(\ref{e.omegakphi4}) and~(\ref{e.Dphi4}) are to be solved
self-consistently.

In order to calculate averages involving on-site momenta we will also need
the LCA renormalization coefficient
\begin{equation}
 \Lambda(t) \equiv \bdave{p_i^2} =
 \frac{3t}{2Q^2RN}\sum_k\sum_{n=-\infty}^\infty
 \frac{f_k^2+\tilde{k}_n}{(\pi n)^2+f_k^2+\tilde{k}_n} ~,
\label{e.Lambdaphi4}
\end{equation}
It is apparent that both $\Lambda$ and $\tilde\mu$ are divergent in the Ohmic
limit, when $\tilde{k}_n=Q\Gamma\pi{n}/(2t)\sim{n}$. This immediately tells us
that the effect of Drude-like dissipation is to enhance the momentum
fluctuations; on the other hand, $D$ decreases when the dissipation strength is
raised, i.e., the coordinate pure-quantum fluctuations are quenched by dissipation.
The physical reason for this lies in the fact that the CL model considers
coordinate coupling with the dissipation bath; for a thorough discussion
see Refs.~\cite{Weiss99,CaldeiraL83}.
The dissipation effects in the thermal averages of quantities depending on both
momenta and coordinates, as internal energy or specific heat, are therefore
unpredictable on a simple basis.

In the nondissipative case the summations over $n$ can be analytically
performed, giving the known results:
\begin{eqnarray}
 D(t;\Gamma{=}0) &=& \frac{Q R}{3N}
 \sum_k\frac1{\Omega_k}\bigg(\coth f_k-\frac1{f_k}\bigg) ~,
\\
 \tilde\mu(t;\Gamma{=}0) &=& \frac{2R}{3N}\sum_k\ln\,\frac{\sinh f_k}{f_k} ~,
\\
 \Lambda(t;\Gamma{=}0) &=& \frac{3}{4QRN}\sum_k\Omega_k\,\coth f_k ~.
\end{eqnarray}

In order to clarify the way the Drude frequency characterizes the
bath bandwidth, we note that the above nondissipative limits are found if
the condition $k_n\ll\omega_k^2$ is satisfied for all values of $n$ and for
all modes $k$. The maximum value of $k_n$, obtained for $n\to\infty$, is
$k_\infty=\gamma\omega_{{}_{\rm{D}}}$. Therefore the $k$-th mode interacts
negligibly with the dissipation bath if
$\Omega_k^2\gg\Gamma\Omega_{{}_{\rm{D}}}$.
Thus the nondissipative limit for the whole system occurs when
$\Gamma\Omega_{{}_{\rm{D}}}\ll\Omega_{k{=}0}^2\sim{1}$. On the other hand, a
crossover should be observed when $\Gamma\Omega_{{}_{\rm{D}}}$
becomes smaller than the squared Debye frequency $\Omega_{k{=}\pi}^2$ so that
the highest-frequency modes become nondissipative. For instance, a less
`dissipative' behavior is to be expected for quantities that depend weakly upon
the low-frequency modes, like the nearest-neighbor distance fluctuation, as we
will show in the next Section. Of course, in the Ohmic limit
$\Omega_{{}_{\rm{D}}}\to\infty$ all modes do dissipate and such a crossover
cannot exist.

We finally observe that the fundamental formula~(\ref{e.aveO})
for thermal averages translates in the present case to
\begin{equation}
 \bave{\hat{\cal{O}}}=
 \Big\langle~\bdave{{\cal{O}}(\bp,\bq{+}\bxi)}\Big\rangle_{\rm{eff}} ~,
\label{e.aveOphi4}
\end{equation}
where the classical-like average with the effective potential reads
\begin{equation}
 \bave{\,\cdot\,}_{\rm{eff}}
 =\frac1{{\cal{Z}}} \bigg(\frac{3t}{4\pi RQ^2}\bigg)^{N/2}
 \int d\bq ~(\,\cdot\,) ~e^{-V_{\rm{eff}}(\bq)/t} ~.
\label{e.aveeff}
\end{equation}

\section{General results}
\label{s.results}

With the above formulation the problem is ready for a numerical evaluation
of Eqs.~(\ref{e.aveOphi4}) and~(\ref{e.aveeff}), that give the partition function
(setting $\hat{\cal{O}}=1$) and the thermal averages of observables. We employed
the numerical transfer matrix technique~\cite{SchneiderS80}, especially useful
in the thermodynamic limit $N\rightarrow\infty$, by which the numerical part
of the calculations is very efficient and can be practically considered exact.
This technique reduces the integrals over the configurations of a one dimensional
array of particles to a secular integral equation, whose evaluation is
implemented numerically using a discrete mesh of points for the possible values
of each degree of freedom. Moreover, the reflection symmetry of the local
effective potential, can be used to halve the dimension of the transfer matrix.

We performed temperature scans over the region of interest and calculated
several quantities, taking the value `per site' for the extensive ones.
From the free energy $f(t)=-N^{-1}\,t\,\ln{\cal{Z}}(t)$ we calculated the
internal energy $u(t)=f(t)-t\partial_tf(t)$ and the specific heat
$c(t)=\partial_tu(t)=-t\partial_t^2f(t)$ by numerical derivation.
In addition we evaluated the thermal average of the squared site coordinate
$\bave{\hat{q}_i^2}$, the local potential $\bave{v(\hat{q}_i)}$,
the square nearest-neighbor displacement $\bave{(\hat{q}_i-\hat{q}_{i-1})^2}$,
and the square momentum $\bave{\hat{p}_i^2}$.
For any parameter set ($t$, $Q$, $R$, $\Gamma$, $\Omega_{{}_{\rm{D}}}$) the
self-consistent computation of $D$ and hence of the last term of $v_{\rm{eff}}(q)$,
Eqs.~(\ref{e.Dphi4}) and~(\ref{e.muphi4}), was performed in negligible computer
time using a continuum termination of the $n$ summation.

\begin{figure}
\centerline{\psfig{bbllx=5mm,bblly=6mm,bburx=184mm,bbury=137mm,%
figure=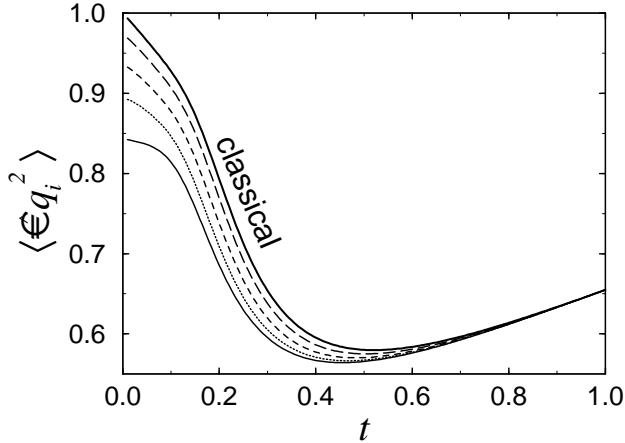,width=82mm,angle=0}}
\caption{Mean-square fluctuations of the coordinate operator
$\langle\hat{q}_i^2\rangle$ vs. reduced temperature $t$, for different
values of the damping strength $\Gamma$, at fixed coupling
parameter $Q=0.2$, kink length $R=5$, and Drude cutoff frequency
$\Omega_{{}_{\rm{D}}}=100$.
Solid line: $\Gamma=0$;
dotted line: $\Gamma=5$;
short-dashed line: $\Gamma=20$;
long-dashed line: $\Gamma=100$;
bold-solid line: classical result.
The latter corresponds also to $\Gamma\to\infty$ in this case.
\label{f.q2med} }
\end{figure}

\begin{figure}
\centerline{\psfig{bbllx=1mm,bblly=4mm,bburx=186mm,bbury=134mm,%
figure=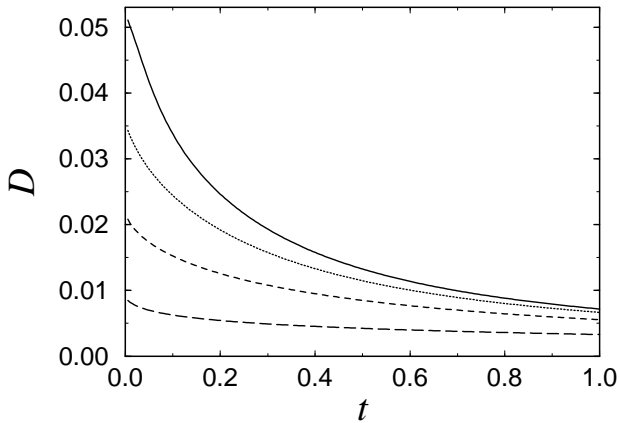,width=82mm,angle=0}}
\caption{Renormalization coefficient $D(t)$ from Eq.~\protect\ref{e.Dphi4},
for different values of the damping strength $\Gamma$.
Parameters and lines as in Fig.~\protect\ref{f.q2med}.
Note that $D(t)\to{0}$ when $\Gamma\to\infty$ .
\label{f.d0} }
\end{figure}

Since the main point of this work is to study the effect of varying the
dissipation parameters, all the quantities reported here are evaluated
for a fixed value of the kink length $R=5$ and using the reference value
$Q=0.2$ for the quantum coupling, which gives fairly strong quantum effects
(the behavior vs. $Q$ and the continuum limit $R\to\infty$, which requires
a careful analysis, are discussed in Ref.~\cite{GTV88fi4}).
For comparison, in the figures we also report the classical results, that
correspond to $Q=0$.
We therefore analyze the dependence upon the damping strength $\Gamma$ and the
Drude frequency $\Omega_{{}_{\rm{D}}}$, which in our scheme fully characterize the
interaction between the system and the environment, i.e., the dissipation.

In order to have a significant dissipative coupling with all the modes
of our system a sufficiently broad environmental coupling spectrum is needed,
i.e., we are interested in values of the Drude frequency $\Omega_{{}_{\rm{D}}}$
such that $\Omega_{{}_{\rm{D}}}\gtrsim\Omega_{k=\pi}^2/\Gamma\sim{4R^2/\Gamma}$.
Taking $\Omega_{{}_{\rm{D}}}=100$ the condition is thus satisfied choosing
values $\Gamma\gtrsim{1}$. In the end of this Section we also consider the effect
of lowering the value of $\Omega_{{}_{\rm{D}}}$ and check the predictions made
at the end of the previous Section.

In Fig.~\ref{f.q2med} we report the temperature behavior of the mean-square
fluctuations of the site coordinate $\langle\hat{q}_i^2\rangle$.
At $t=0$, in the classical case the coordinate lies in the minima $q^2=1$
of the potential (\ref{e.vphi4}), while in the quantum non-dissipative case
the value at $t=0$ is smaller, $\langle\hat{q}_i^2\rangle=1-3D\simeq{0.84}$,
due to the quantum fluctuations which occur more likely towards the barrier
rather than the steeper walls, and corresponds to the minima of $v_{\rm{eff}}(q)$.
The classical thermal fluctuations enhance the same effect and
$\langle\hat{q}_i^2\rangle$ decreases at finite temperature, until when at
$t\gtrsim{0.5}$ kinks are excited in the system ($t$ being the temperature
in kink-energy units) and the coordinate distribution begins to extend towards
the walls, so that $\langle\hat{q}_i^2\rangle$ starts to increase. Finally,
in the high-temperature limit all curves collapse into the classical one.
Switching on the damping strength $\Gamma$ the average of $\hat{q}_i^2$
tends to come back to the classical value: in other words, the pure-quantum
fluctuations of the coordinates are `effectively' quenched by dissipation,
as we already remarked and can be observed in Fig.~\ref{f.d0}, where
the pure-quantum renormalization coefficient $D(t)=\dave{\xi_i^2}$
of Eq.~(\ref{e.Dphi4}) is reported for different values of $\Gamma$.

In Fig.~\ref{f.q2med} one can also observe that before
collapsing into one single curve as $t$ is raised, the finite-$\Gamma$
curves get closer to the quantum result. This can be explained by considering
the expression~(\ref{e.Dphi4}) of the renormalization coefficient $D(t)$
together with Eq.~(\ref{e.knphi4}): a rough estimate tells that
dissipation is not effective (in quenching the value of $D$) when
$\Gamma\ll\Gamma_{\rm{c}}(t)=2t\pi/{Q}$; for instance,
$\Gamma_{\rm{c}}(t{=}0.2)\sim{6}$ and $\Gamma_{\rm{c}}(t{=}0.5)\sim{16}$,
explaining the behavior of the curve for $\Gamma=5$.
Since rising the temperature drives the system towards the classical behavior,
this phenomenon is in agreement with the concept that the thermodynamics of a
classical system is not affected by dissipation.

\begin{figure}
\centerline{\psfig{bbllx=21mm,bblly=9mm,bburx=184mm,bbury=134mm,%
figure=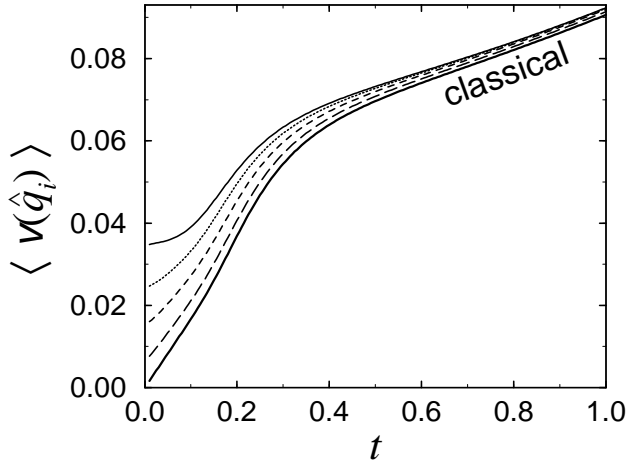,width=82mm,angle=0}}
\caption{Average local potential energy $\langle{v}(\hat{q}_i)\rangle$
vs. reduced temperature $t$, for different values of the damping
strength $\Gamma$. Parameters and lines as in Fig.~\protect\ref{f.q2med}.
Note that $\langle{v}(\hat{q}_i)\rangle$ tends to
the classical behavior for $\Gamma\to\infty$.
\label{f.vmed} }
\end{figure}

On the same basis one can interpret the behavior of the average local
potential $\langle{v}(\hat{q}_i)\rangle$, which is reported in Fig.~\ref{f.vmed}.
It raises with temperature because of thermal and quantum fluctuations and
in the classical case at low $t$ the initial slope is
$R/(3\sqrt{1{+}4R^2})\sim{1/6}$; when $t\gtrsim 0.5$ the nonlinearity
dominates and the curves show a crossover to a smaller slope.
Since the potential depends on the coordinates only, the inclusion of
dissipation leads the quantum behavior back towards the classical curve.

On the other hand, the fluctuations of the momenta are enhanced by
dissipation, and the role of the damping effects is non predictable on
a simple basis if one considers thermodynamic quantities where both
coordinates and momenta enter into play. One of these is the specific
heat $c(t)$, which is proportional to the mean-square fluctuation of the
Hamiltonian:
\begin{equation}
 c(t)=(\varepsilon_{{}_{\rm{K}}}t)^{-2} ~\Big\langle
 \big(\hat{\cal{H}}-\big\langle\hat{\cal{H}}\big\rangle\big)^2 \Big\rangle ~.
\end{equation}

\begin{figure}
\centerline{\psfig{bbllx=6mm,bblly=14mm,bburx=184mm,bbury=141mm,%
figure=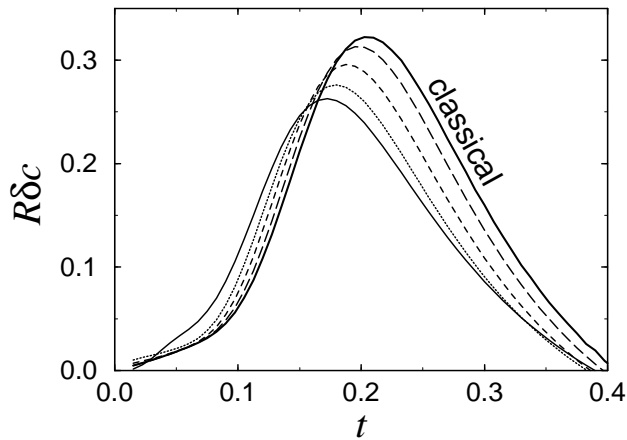,width=82mm,angle=0}}
\caption{Nonlinear contribution to the specific heat, defined as
$\delta{c}(t)=c(t)-c_{\rm{h}}(t)$ (times the kink length $R$),
vs. reduced temperature $t$, for different values of the damping strength
$\Gamma$. Parameters and lines as in Fig.~\protect\ref{f.q2med}.
For $\Gamma\to\infty$, $\delta{c}(t)$ tends to the classical behavior.
\label{f.cnl} }
\end{figure}

In Fig.~\ref{f.cnl} we first report the temperature behavior of the
{\em nonlinear} part $\delta{c}(t)$ of the specific heat, namely its
total value minus the corresponding (dissipative) harmonic contribution,
$\delta{c}(t)=c(t)-c_{\rm{h}}(t)$.
This quantity is very sensitive to the nonlinearity of the system,
since its value is zero in a harmonic approximation, and the fact that
the PQSCHA retains all classical nonlinear features is crucial for
getting $\delta{c}$.
Increasing the strength of the environmental coupling $\Gamma$ the curves
can be seen to tend to the classical result, giving us a nontrivial
outcome of the formalism.

\begin{figure}
\centerline{\psfig{bbllx=10mm,bblly=14mm,bburx=184mm,bbury=144mm,%
figure=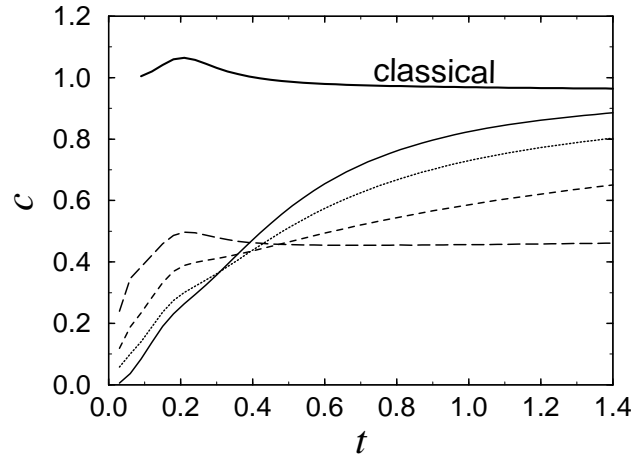,width=82mm,angle=0}}
\caption{Total specific heat $c(t)$ vs. reduced temperature $t$,
for different values of the damping strength $\Gamma$.
Parameters and lines as in Fig.~\protect\ref{f.q2med}.
Note that for $\Gamma\to\infty$, $c(t)$ tends to $c_{\rm{cl}}-1/2$.
\label{f.ctot} }
\end{figure}

\begin{figure}
\centerline{\psfig{bbllx=19mm,bblly=9mm,bburx=184mm,bbury=136mm,%
figure=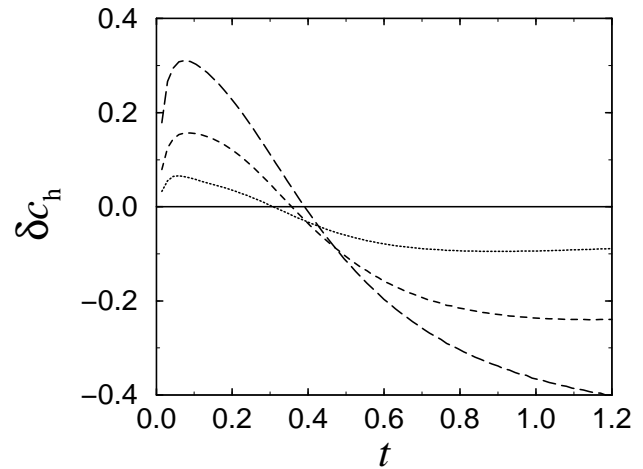,width=82mm,angle=0}}
\caption{Effect of the dissipation strength on the harmonic
contribution to the specific heat
$\delta{c}_{\rm{h}}(t,\Gamma)=c_{\rm{h}}(t,\Gamma)-c_{\rm{h}}(t,0)$
vs. reduced temperature $t$, for different values of the damping
strength $\Gamma$.
Parameters and lines as in Fig.~\protect\ref{f.q2med}.
\label{f.ch} }
\end{figure}

This is remarkable, also in view of the fact that the total specific heat,
shown in Fig.~\ref{f.ctot}, does not behave like its nonlinear part.
Increasing $\Gamma$ the curves do not tend to the classical one: rather,
the classical limit is more rapidly reached in the non dissipative case,
$\Gamma=0$. In particular, one can observe in this figure two significative
features:
~{\em(i)}~the strong dependence on $\Gamma$, with the curves crossing at
$t\sim{0.4}$, and that
~{\em(ii)}~at large values of $\Gamma$, the specific heat seems to tend to
the classical result minus one half, $c(t)\to{c}_{\rm cl}(t)-1/2$.
These features arise from two different effects.

The first one can be explained by the consequences of dissipation on the
linear contribution to the specific heat.
In Fig.~\ref{f.ch} we report indeed the difference
between the dissipative and the non dissipative harmonic specific heat,
$\delta{c}_{\rm{h}}(t,\Gamma)=c_{\rm{h}}(t,\Gamma)-c_{\rm{h}}(t,0)$, where
the crossing of the curves at $t\sim{0.4}$ can be observed: the dissipation
bath appears to absorb energy from the system at low $t$ and to give it back
when $t\gtrsim{0.4}$. This is therefore a purely harmonic effect, whose origin
can be understood thinking of the hybridization of two oscillators of equal
frequency $\omega$, one representing the system and the other the bath: the
dissipation arises indeed from such hybridizations, whose effect is maximized
at resonance. Adding a small linear coordinate coupling $2\gamma\omega{q_1}q_2$
and diagonalizing one gets two diagonal frequencies,
$\omega_\pm=\omega\pm\gamma$, and in fact it can be seen that the quantity
\begin{equation}
 \delta{c}_{\rm{h}}(t,\gamma)=
 c_{\rm{h}}(\omega{+}\gamma)+c_{\rm{h}}(\omega{-}\gamma)-2c_{\rm{h}}(\omega)
\end{equation}
with $c_{\rm{h}}(\omega)=(f/\sinh{f})^2$, $f=\omega/2t$,
behaves qualitatively like $\delta{c}_{\rm{h}}(t,\Gamma)$ in Fig.~\ref{f.ch}.

\begin{figure}
\centerline{\psfig{bbllx=10mm,bblly=24mm,bburx=189mm,bbury=233mm,%
figure=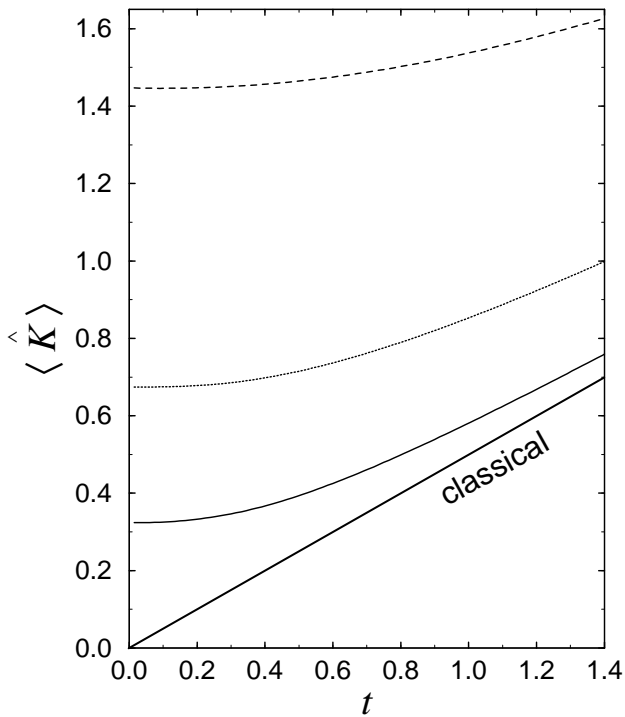,width=82mm,angle=0}}
\caption{Average kinetic energy $\langle\hat{K}\rangle$ vs. reduced
temperature $t$, for different values of damping intensity $\Gamma$.
Parameters and lines as in Fig.~\protect\ref{f.q2med}.
One can see that $\langle\hat{K}\rangle$ diverges for $\Gamma\to\infty$,
while it becomes flatter and flatter.
\label{f.kin} }
\end{figure}

\begin{figure}
\centerline{\psfig{bbllx=8mm,bblly=23mm,bburx=199mm,bbury=262mm,%
figure=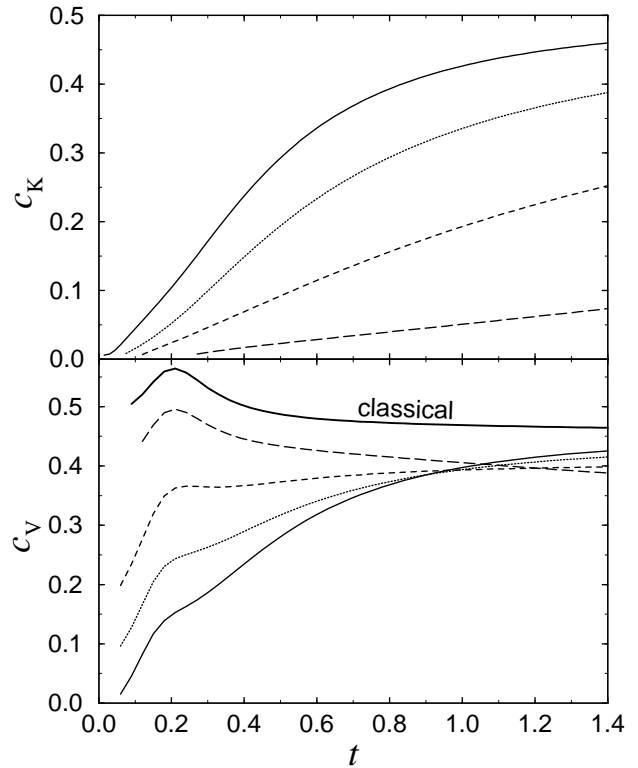,width=82mm,angle=0}}
\caption{Kinetic and interaction parts of the specific heat,
namely $c_{{}_{\rm{K}}}(t)=\partial_t\langle\hat{K}\rangle$ and
$c_{{}_{\rm{V}}}(t)=\partial_t\langle\hat{V}\rangle$, vs. reduced
temperature $t$, for different values of damping intensity $\Gamma$.
Parameters and lines as in Fig.~\protect\ref{f.q2med}.
Note that $c_{{}_{\rm{K}}}(t)\to{0}$ for $\Gamma\to\infty$.
\label{f.ckcv} }
\end{figure}

In order to understand the second feature of the total specific heat, we
report in Fig.~\ref{f.kin} the thermal average of the kinetic energy per site
$\langle\hat{K}\rangle=(Q^2R/3)\,\langle\hat{p}_i^2\rangle$.
We recall that increasing $\Gamma$ the quantum fluctuations of momenta are
enhanced, as it can be seen indeed in Fig.~\ref{f.kin} where the curve that
reaches more rapidly the classical limit is the non dissipative one, as it
also occurs for the specific heat in Fig.~\ref{f.ctot}. However, while
$\langle\hat{K}\rangle$ increases with damping, its slope decreases, as
shown in the upper part of Fig.~\ref{f.ckcv}, where we compare the contributions
to the specific heat arising from the kinetic and the interaction parts of the
Hamiltonian, namely $c_{{}_{\rm{K}}}(t)=\partial_t\langle\hat{K}(t)\rangle$ and
$c_{{}_{\rm{V}}}(t)=\partial_t\langle\hat{V}(t)\rangle$.
When $\Gamma$ increases $c_{{}_{\rm{K}}}(t)$ raises more and more slowly
towards the classical value $1/2$, as if the kinetic specific heat were
quenched by dissipation, while $c_{{}_{\rm{V}}}(t)$ tends to the classical curve.
The combined effect is therefore that for large $\Gamma$ the total specific heat
(Fig.~\ref{f.ctot}) tends to its classical interaction part and seems to be
lacking the kinetic contribution.

The condition for $\langle\hat{K}\rangle$ and $c_{{}_{\rm{K}}}(t)$ to reach
their classical limits ($\sim{t/2}$ and $\sim{1/2}$, respectively) is easily
found by writing, from Eq.~(\ref{e.Lambdaphi4}), the thermal average of the
kinetic energy as
\begin{equation}
 \langle\hat{K}\rangle=\frac t2 + \frac tN \sum_k \sum_{n=1}^{\infty}
 {\frac{f_k^2+\tilde{k}_n}{(\pi n)^2+f_k^2+\tilde{k}_n}} ~,
\end{equation}
and using Eqs.~(\ref{e.knphi4}) and~(\ref{e.omegakphi4}). At the end the
following condition is obtained
\begin{equation}
 t\gg \frac{Q}{2\sqrt{3}}\sqrt{2R^2+\Gamma\Omega_{{}_{\rm{D}}}} ~,
\end{equation}
which for the values $R{=}5$, $Q{=}0.2$, and $\Omega_{{}_{\rm{D}}}{=}100$
considered here amounts to $t\gg{0.6}\sqrt{0.5{+}\Gamma}$, which is
well verified from our numerical outcomes.

\begin{figure}
\centerline{\psfig{bbllx=16mm,bblly=15mm,bburx=187mm,bbury=251mm,%
figure=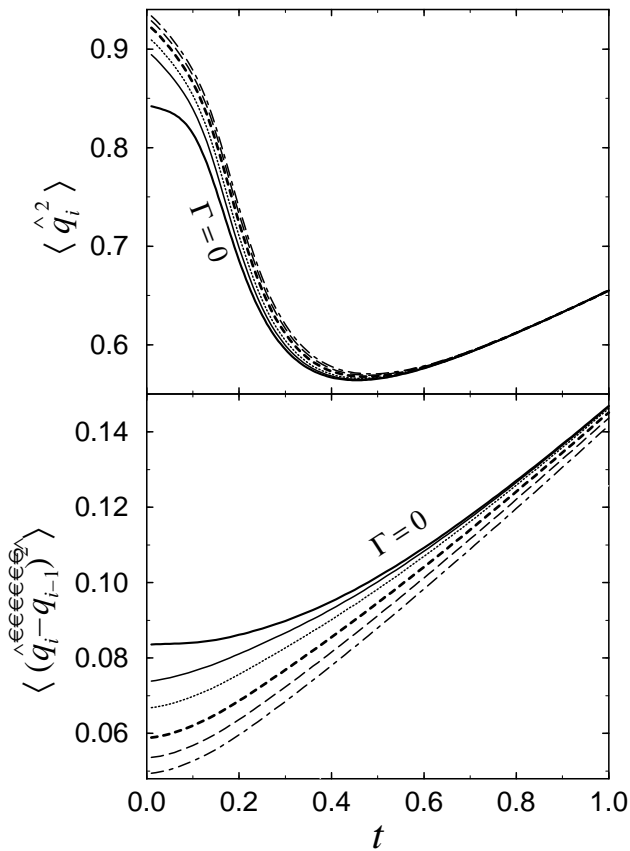,width=82mm,angle=0}}
\caption{Thermal averages $\langle\hat{q}_i^2\rangle$ and
$\langle(\hat{q}_i{-}\hat{q}_{i-1})^2\rangle$, vs. reduced temperature $t$, for
different values of the Drude frequency $\Omega_{{}_{\rm{D}}}$,
at fixed coupling $Q=0.2$, kink length $R=5$, and damping strength $\Gamma=20$.
Solid line: $\Omega_{{}_{\rm{D}}}=1$;
dotted line: $\Omega_{{}_{\rm{D}}}=3$;
short-dashed line: $\Omega_{{}_{\rm{D}}}=10$;
long-dashed line: $\Omega_{{}_{\rm{D}}}=30$;
dot-dashed line: $\Omega_{{}_{\rm{D}}}=\infty$ (Ohmic limit);
bold-solid line: nondissipative result, i.e., $\Gamma=0$ or
$\Omega_{{}_{\rm{D}}}=0$.
\label{f.q2medij} }
\end{figure}

Finally, let us discuss the role of the cutoff frequency
$\Omega_{{}_{\rm{D}}}$. As we have seen before, the mechanism of the coupling
between the system and the environment can be traced back, in a simplified
scheme, to the hybridization of oscillators. This process is effective only if
the frequencies of the oscillators are close to resonance. So the modes
$\{\Omega_k\}$ of the system, which couple with the environment, are those
which do not exceed the cutoff frequency $\sqrt{\Gamma\Omega_{{}_{\rm{D}}}}$,
as discussed at the end of Section~\ref{ss.phi4.veff}. This gives rise to
interesting effects: as an instance, in Fig.~\ref{f.q2medij} we analyze the
effect of varying $\Omega_{{}_{\rm{D}}}$ on the averages of the on-site
coordinate $\big\langle\hat{q}_i^2\big\rangle$ and of the nearest-neighbor
displacement
\begin{equation}
 \big\langle(\hat{q}_i-\hat{q}_{i-1})^2\big\rangle=
 \Big\langle (q_i-q_{i-1})^2 \Big\rangle_{\rm{eff}} + {\cal{D}}(t) ~.
\end{equation}
The latter involves the nearest-neighbor renormalization coefficient
${\cal{D}}=2(C_{ii}-C_{i,i-1})$, which is expressed as
\begin{equation}
 {\cal{D}}(t)=\frac{Q^2R}{3tN}\sum_k\sum_{n=1}^\infty
 \frac{2(1-\cos{k})}{(\pi n)^2+f_k^2+\tilde{k}_n} ~.
\label{e.DDphi4}
\end{equation}
Comparing with Eq.~(\ref{e.Dphi4}) it appears that the contribution from the
low-frequency modes is less relevant due to the $k$-dependent factor, allowing
one to expect that when the cutoff frequency becomes smaller than the Debye
frequency $\Omega_{k{=}\pi}\sim{2R}$, i.e., when
\begin{equation}
 \Omega_{{}_{\rm{D}}}\lesssim 4R^2/\Gamma ~,
\label{e.condOmegaD}
\end{equation}
the effect of dissipation on
$\big\langle(\hat{q}_i-\hat{q}_{i-1})^2\big\rangle$ should decrease more
pronouncedly than for $\big\langle\hat{q}_i^2\big\rangle$.
This is indeed apparent from Fig.~\ref{f.q2medij}, where the damping strength
is set to $\Gamma{=}20$ and the above condition reads
$\Omega_{{}_{\rm{D}}}\lesssim{5}$.

\section{Conclusions}

In this paper we have derived the {\em pure-quantum self-consistent approximation}
(PQSCHA) formalism for an interacting many-body system with Caldeira-Leggett
dissipation and nondiagonal quadratic kinetic energy. The PQSCHA allows one to
reduce quantum mechanical thermodynamic calculations to a classical-like
configuration integral, where in the present case also the quantum effects of
dissipation are included. In order to deal with many degrees of freedom the
necessary {\em low-coupling} approximation has been introduced. The latter,
if the system's symmetries are exploited, results in very simple expressions
for the renormalization coefficients appearing in the theory. This is shown
in detail for the case of translation symmetry.

The rest of the paper is devoted to the application of the framework to
the discrete $\phi^4$ one-dimensional field, whose strong nonlinearity
results in the characteristic kink excitations which play a determinant role
in the thermodynamics. A Drude-like spectrum of the environmental coupling
is chosen for dissipation, and its influence on several quantities is
analyzed when the two characterizing parameters, i.e., dissipation strength and
bandwidth, are varied.
The PQSCHA is a unique tool for dealing with such a system, since any
approximate theory must retain the strong nonlinearity, which has a mainly
classical character, and this rules out conventional perturbative approaches.
In general, the inclusion of dissipation through coordinate coupling with the
environment results in quenched quantum fluctuations of the coordinates, while
those of the momenta are emphasized.
Interesting and nontrivial behavior is found for the specific heat, since
the prevalence of either of the mentioned effects is not predictable
on simple grounds.

The method proves to be very useful for studying our model system, and we
believe it will find several application in physical contexts where both
quantum and dissipative effects play an important role.

\appendix
\section{The trial density matrix}
\label{a.rho0}

In this appendix we report the details of the calculations which lead to
the results reported in Section~\ref{ss.PQSCHA}.

Inserting the proper constraints in Eq.~(\ref{e.brho0eps}) and, using the
explicit form (\ref{e.S0Matsu}) of the trial action in terms of Matsubara variables,
one has
\begin{eqnarray}
 & & \bar\rho_0(\bq'',\bq';\bar\bq) =
 \lim\limits_{\varepsilon\to 0} \,\frac{e^{-\beta\,w(\bar\bq)}}{{\cal F}_\varepsilon}
 \oint {\cal{D}}\big[\bq(u)\big]
\nonumber\\
 & & ~~~~~~~~\times
 \,\delta\big(\bar\bq{-}\bar\bq\big[q(u)\big]\big)
 \,\delta\big(\bq(\varepsilon){-}\bq'\big)
 \,\delta\big(\bq(\beta{-}\varepsilon){-}\bq''\big)
\nonumber\\
 & & ~~~~~~~~\times
 \,e^{-\beta ({}^{\rm{t}}\!\bx_n \bPhi_n \bx_n
 +{}^{\rm{t}}\!\by_n \bPhi_n \by_n)} \,,
\label{e.brho01}
\end{eqnarray}
where the path-integral measure is given by Eq.~(\ref{e.measure}).
Representing the delta functions that fix $\bq'$ and $\bq''$ as
$\delta(\bq)= (2\pi)^{-N}\int{d\bv}\,{}~ e^{-i\,{}^{\rm{t}}\!\bq \bv}$\,,
taking the explicit expressions of $\bq(\varepsilon)$ and of
$\bq(\beta-\varepsilon)$ from Eq.~(\ref{e.qn}),
and using the general Gaussian quadrature formula
\begin{equation}
 \int d\bx ~e^{-\frac12\,{}^{\rm{t}}\!\bx\bPhi\bx+i{}^{\rm{t}}\!\bx\bq}
 = \frac{(2\pi)^{N/2}}{\sqrt{\det\bPhi}}
 ~e^{-\frac12\,{}^{\rm{t}}\!\bq \bPhi^{-1} \bq} ~,
\label{e.Gaussint}
\end{equation}
one can integrate over the variables $\bx_n$ and $\by_n$, obtaining
\begin{eqnarray}
 & & \bar\rho_0(\bq'',\bq';\bar\bq) =
 \bigg(\frac{m}{2\pi\beta}\bigg)^{N/2}
 \,\frac{e^{-\beta\,w(\bar\bq)}}{\mu(\bar\bq)}
 \lim\limits_{\varepsilon\to 0}
 \,\frac1{{\cal F}_\varepsilon}
 \int \frac{d\bv^+d\bv^-}{(2\pi)^{2N}}
\nonumber\\
 & &~~~~~~ \times
 \,e^{-\frac12\big({}^{\rm{t}}\!\bv^+ \bC_\varepsilon \bv^+
 + {}^{\rm{t}}\!\bv^- \bS_\varepsilon \bv^-\big)}
 ~e^{i[{}^{\rm{t}}\!\bv^+\bxi+{}^{\rm{t}}\!\bv^-\bzeta]}
\label{e.brho02}
\end{eqnarray}
where $\bxi=\frac12(\bq''{+}\bq')-\bar\bq$ and $\bzeta=\bq''-\bq'$,
$\bv^+=\bv''+\bv'$ and $\bv^-=\frac12(\bv''-\bv')$; furthermore, the function
$\mu(\bar\bq)$ is defined as
\begin{equation}
 \mu(\bq) = \sum\limits_{n=1}^{\infty}\,
 \ln\,\frac{\det\bPhi_n(\bq)}{(m\nu_n^2)^N}~,
\end{equation}
and
\begin{eqnarray}
 \bC_\varepsilon &=&
 \frac2\beta\sum_{n=1}^\infty \bPhi_n^{-1} \cos^2\nu_n\varepsilon ~,
\label{e.Cmatr}
\\
 \bS_\varepsilon &=&
 \frac8\beta\sum_{n=1}^\infty \bPhi_n^{-1} \sin^2\nu_n\varepsilon ~.
\label{e.Smatr}
\end{eqnarray}
Using again Eq.~(\ref{e.Gaussint}) the reduced density can then be written as
\begin{eqnarray}
 & & \bar\rho_0(\bq'',\bq';\bar\bq) =
 \bigg(\frac{m}{2\pi\beta}\bigg)^{N/2}
 \,\frac{e^{-\beta\,w(\bar\bq)}}{\mu(\bar\bq)}
 \lim\limits_{\varepsilon\to 0}
 \,\frac1{{\cal F}_\varepsilon}
\nonumber\\
 & & ~~\times
 \,\frac1{(2\pi)^N\sqrt{\det\bC_\varepsilon\det\bS_\varepsilon}}
 ~e^{-\frac12\big({}^{\rm{t}}\!\bxi \bC_\varepsilon^{-1} \bxi
 + {}^{\rm{t}}\!\bzeta \bS_\varepsilon^{-1} \bzeta\big)} ~.
\label{e.brho03}
\end{eqnarray}
In order to perform the limit, we have to expand $\bS_\varepsilon$ for
small $\varepsilon$. Let us first consider the associated matrix
\begin{equation}
 \widetilde\bS_\varepsilon=\bA^{-1}\bS_\varepsilon\bA^{-1}
 =\frac8\beta\sum_{n=1}^\infty\big[\nu_n^2+\bTheta_n\big]^{-1}
 \sin^2\nu_n\varepsilon~,
\end{equation}
with $\bTheta_n=\bA\big[\bB^2+\bK_n\big]\bA$. Subtracting from the exactly
summable series~\cite{GradshteynR80}
\begin{equation}
 g_\varepsilon
 \equiv \frac8\beta\sum_{n=1}^\infty\frac{\sin^2\nu_n\varepsilon}{\nu_n^2}
 = 2\varepsilon\bigg(1-2\frac\varepsilon\beta\bigg) ~,
\label{e.g-eps}
\end{equation}
that is verified for $|\varepsilon|\leq\beta/2$, we have
\begin{equation}
 g_\varepsilon-\widetilde\bS_\varepsilon
 =\frac8\beta\sum_{n=1}^\infty \bTheta_n\big[\nu_n^2+\bTheta_n\big]^{-1}
 \frac{\sin^2\nu_n\varepsilon}{\nu_n^2}~.
\end{equation}
Provided that $\bTheta_n/n^\alpha\to{0}$ for $n\to\infty$ and for some
$\alpha<1$ this series and its first and second derivatives are uniformly
convergent: therefore the limit $\varepsilon\to{0}$ can be taken under the
summation yielding the Taylor expansion
\begin{equation}
 g_\varepsilon-\widetilde\bS_\varepsilon
 =\varepsilon^2 ~\frac8\beta\sum_{n=1}^\infty
 \bTheta_n\big[\nu_n^2+\bTheta_n\big]^{-1} +O(\varepsilon^3)
\end{equation}
and finally giving
$\widetilde\bS_\varepsilon=2\varepsilon(1-2\varepsilon\widetilde\bLambda)
+O(\varepsilon^3)$~, with
\begin{equation}
 \widetilde\bLambda=\frac1\beta\sum_{n=-\infty}^\infty
 \bTheta_n\big[\nu_n^2+\bTheta_n\big]^{-1} ~.
\end{equation}
Then, we also have $\det\bS_\varepsilon\simeq(2\varepsilon/m)^N$
and $\bS_\varepsilon^{-1}\simeq(\bA^{-2}+2\varepsilon\bLambda)/2\varepsilon$,
with $\bLambda=\bA^{-1}\widetilde\bLambda\bA^{-1}$, and in view of
Eq.~(\ref{e.F-eps}) the limit for $\varepsilon\to{0}$ in
Eq.~(\ref{e.brho03}) can be evaluated, leaving
\begin{eqnarray}
 & & \bar\rho_0(\bq'',\bq';\bar\bq) =
 \bigg(\frac{m}{2\pi\beta}\bigg)^{N/2}
 e^{-\beta\,w(\bar\bq)-\mu(\bar\bq)}
\nonumber\\
 & & ~~~~~~~~~~\times
 \frac1{\sqrt{(2\pi)^N\det\bC}}
 ~e^{-\frac12({}^{\rm{t}}\!\bxi \bC^{-1} \bxi
 + {}^{\rm{t}}\!\bzeta \bLambda \bzeta)} ~.
\label{e.brho04}
\end{eqnarray}
Going over to the Weyl symbol~\cite{Berezin80,HilleryCSW84} corresponding
to the reduced density matrix, one obtains
\begin{eqnarray}
 \bar\rho_0(\bp,\bq;\bar\bq) &=& (2\pi)^N
 \bigg(\frac{m}{2\pi\beta}\bigg)^{N/2}
 e^{-\beta\,w(\bar\bq)-\mu(\bar\bq)}
\nonumber\\
 & & \times
 \,\frac{e^{-\frac12\,{}^{\rm{t}}\!\bxi \bC^{-1} \bxi}}{\sqrt{(2\pi)^N\det\bC}}
 \,\frac{e^{\frac12\,{}^{\rm{t}}\!\bp \bLambda^{-1} \bp}}{\sqrt{(2\pi)^N\det\bLambda}}
 ~.
\label{e.brho0pq}
\end{eqnarray}
This Gaussian distribution for $\bxi\equiv\bq{-}\bar\bq$ and $\bp$ determines
exactly the double-bracket average of Eqs.~(\ref{e.Cij}) and~(\ref{e.Lambdaij}),
and it is then easy to derive Eq.~(\ref{e.aveO}).



\end{document}